\documentclass[preprint,showpacs,preprintnumbers,amsmath,amssymb,superscriptaddress]{revtex4}

\include{ams}
\usepackage{amsmath,amssymb,amsthm}
\usepackage{dcolumn}
\usepackage{bm}
\usepackage{graphicx}

\newcommand{\RR}{\mathbf{r}}

\newcommand{\bb}{\mathbf{b}}

\begin{document}

\title{Optimized Effective Potentials in Finite Basis Sets}

\author{Tim Heaton-Burgess}
\author{Felipe A. Bulat}
\author{Weitao Yang}
\affiliation{Department of Chemistry, Duke University, Durham, North
Carolina 27708, USA}


\date{\today}

\begin{abstract}
The finite basis optimized effective potential (OEP) method within
density functional theory is examined as an ill-posed problem. It is
shown that the generation of nonphysical potentials is a
controllable manifestation of the use of unbalanced, and thus
unsuitable, basis sets. 
 A modified functional incorporating a regularizing smoothness measure of
the OEP is introduced. This provides a condition on balanced basis
sets for the potential, as well as a method to determine the most
appropriate OEP potential and energy from calculations performed
with any finite basis set.


\end{abstract}

\maketitle

Kohn-Sham density functional theory (KS DFT) \cite{parr89} enjoys
wide application owing to the computational accessibility afforded
by formulating the many body problem in terms of the noninteracting
KS reference system. However, the exact exchange-correlation energy
functional $ E_{xc}$ and corresponding local potential,
$v_{xc}$, are unknown. The future success of DFT  is dependent on
the availability of suitable approximations for $ E_{xc}$.
Significant interest is being shown for the development of implicit
density functionals, depending explicitly on the KS orbitals
\cite{mori06}. For such functionals, $v_{xc}$ cannot be directly
obtained as a simple functional derivative and requires an OEP
method for its determination
\cite{hirata99,gorling99,YANG_WU_PRL_02}.


The OEP concept first appeared within the Hartree-Fock (HF)
formalism 
\cite{talman76}, and later was employed in DFT \cite{sahni82} using
the HF exact exchange energy functional (EXX). These conventional
approaches identify the local KS potential for exact exchange as
that obtained from solving $\delta E[\{\phi_{i}^{\textrm{KS}}\}] /
\delta v_{s}(\RR) = 0$, which in turn leads to a linear integral
equation to be solved for the OEP. The rigorous justification of the
OEP method as a variational minimization has however only recently
been provided within the potential functional formulation
\cite{YANG_AYERS_WU_PRL_92}. This formulation lends itself to a
direct approach to solving for the OEP \cite{YANG_WU_PRL_02} where
the minimization of the energy functional over local potentials is
considered.

The discrete representation of the OEP problem in finite basis sets
for both the Kohn-Sham orbitals and the potential can be
 ill-posed  \cite{hirata01,staroverov06}: While the total
OEP energy is  stable with respect to the changes in the potential,
there can be many different potentials, including nonphysical ones,
having numerically degenerate total energies.
%
%
%
This has lead to some degree of confusion of what in fact
constitutes a valid, finite basis, OEP implementation
\cite{staroverov06}, when the ill-posedness is not appropriately
accounted for.


In this Letter, we will show that the ill-posed nature of the
discrete OEP originates from the use of basis sets that are
unsuitable, as a result of being unbalanced. This ill-posedness does
not imply that the OEP method is unphysical by construction, rather,
just that it can lead to nonphysical potentials.
The regularization method developed in this work insures that the
physical context of finite basis OEP calculations are maintained in
all cases, with generation of physically meaningful potentials.

The following analysis reveals the physical origin of the ill-posed
nature in the discrete OEP problem. In determining the OEP, one
looks for the total-energy minimizing potential $v_{s}(\mathbf{r})$
whose lowest $N$ eigenstates are the set of wavefunctions with the
finite basis set expansion,
$|\phi_{i}\rangle=\sum_{\mu}c_{i\mu}|\chi_{\mu}\rangle$, for
$i=1,...,N$, satisfying the one-electron equations
$\sum_{\mu}\langle\chi_{\nu}|-\frac{1}{2}\nabla^{2}+v_{s}(\mathbf{r})|\chi_{\mu}\rangle
c_{i\mu}=\epsilon_{i}\sum_{\mu}\langle\chi_{\nu}|\chi_{\mu}\rangle
c_{i\mu}. \label{eq:discretOEP}$
This equation is the only link from $v_{s}(\mathbf{r})$ to the OEP
energy. If a variation of $v_{s}(\mathbf{r})$ is outside of the
range of the orbital basis set $\{\chi_{\mu}\}$ -- meaning such a
variation in the potential leads to numerically the same total
energy and orbitals $|\phi_{i}\rangle $
-- then this variation can lead to a nonphysical potential. The
nonphysical potentials reported in Ref. \cite{staroverov06} appear
to come from such a construction. For nonphysical potentials with
wild oscillations, the orbital basis set is certainly inadequate for
solving the one-electron equations and the discrete OEP equation is
a poor approximation to the original OEP problem. Given an orbital
basis set $\{\chi_{\mu}\}$, there are many variations of
$v_{s}(\mathbf{r})$ that can lead to such behavior, one example
being adding to $v_{s}(\mathbf{r})$ Gaussian functions with
exponents much higher than the exponents in the orbital basis set
$\{\chi_{\mu}\}$. We consider such a potential basis set
\emph{unbalanced} with the orbital basis set.

While a careful choice of basis sets for the orbitals and the
potential can lead to accurate potentials, as shown in many previous
calculations \cite{gorling99,hirata01,YANG_WU_PRL_02}, we aim here
to develop a robust method which can deliver accurate OEP potentials
and energies, with both balanced and unbalanced basis sets.

To examine the reliability of our approach we use the OEP procedure
with the LDA (SVWN5) functional so that a direct comparison between
the calculated OEP potentials and the LDA potential can be made. The
success of our method for LDA OEP transfers to EXX as will be shown.

Our OEP implementation is the direct optimization approach of Yang
and Wu \cite{YANG_WU_PRL_02} where the trial potential is expanded
in a finite basis set, $\{g_{t}\}$, as
$v_{s}^{\sigma}(\RR)=v_{\textrm{ext}}(\RR)+v_{0}(\RR)+\sum_{t}b_{t}^{\sigma}g_{t}(\RR)$.
Here $v_{\textrm{ext}}$ is the external potential of the system
under consideration and $v_{0}$ is a fixed reference potential,
 taken as the Fermi-Amaldi potential (or the Coulomb
potential for LDA) for the sum of the atomic densities so to enforce
the correct asymptotic behavior upon $v_{s}$. Transferring the
functional dependence from the KS potential on to the expansion
coefficients $\{b_{t}^{\sigma}\}$ in this way gives rise to an
efficient implementation of the OEP  based on the unconstrained
minimization of $E(\{b_{t}^{\sigma}\})$ with readily available
analytic derivatives \cite{YANG_WU_PRL_02}.
The iterative minimization of $E(\{b_{t}^{\sigma}\})$ is efficiently
achieved using an approximate Newton method \cite{WU_YANG_JTCC_03}
which takes an approximate form for the Hessian as
$H_{ut}^{\sigma}=\sum_{ia}\frac{\langle\phi_{a\sigma}|g_{u}|\phi_{i\sigma}\rangle\langle\phi_{a\sigma}|g_{t}|\phi_{i\sigma}\rangle}{\epsilon_{i\sigma}-\epsilon_{a\sigma}}.$


In the following we will consider two forms of regularization for
extracting physical potentials from ill-posed finite basis  OEP
calculations. The first, used by us before
\cite{YANG_WU_PRL_02,WU_YANG_JTCC_03}, however will be shown to be
inadequate to remove \emph{all} irregularities in the resulting
potential. We then introduce a new form of regularization for the
OEP that produces potentials of the highest quality, along with the
corresponding energy.

We will focus on the specific example of argon with the orbital
basis cc-pVDZ and consider the effects of different potential basis
on the potentials and energies obtained from the OEP procedure. The
specific basis sets we use for the potential are the orbital basis
itself, and three s-type even tempered (ET) basis sets denoted as
ArX (X=64,1024,8192) with exponents $2^{n}, -4\leq n\leq
\log_{2}(\textrm{X})$.

\begin{figure}[tbh]
\centering{\includegraphics[height=8cm]{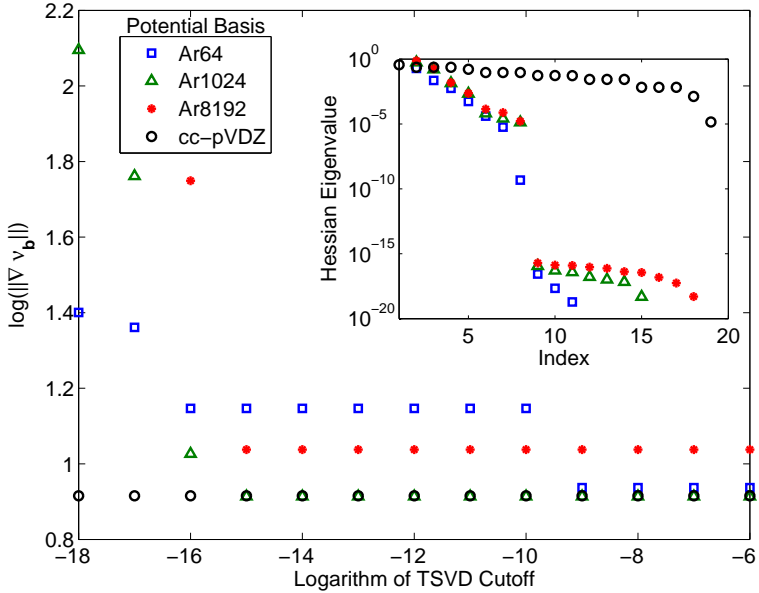}}
\caption{Variation in the smoothness of the LDA OEP potential as a
function of the SVD cutoff for argon (orbital basis cc-pVDZ) with
four different potential basis sets. Insert: Spectrum of the
approximate
Hessian 
for each basis set.}
\end{figure}

\begin{figure}[htb]
\centering{\includegraphics[height=8cm]{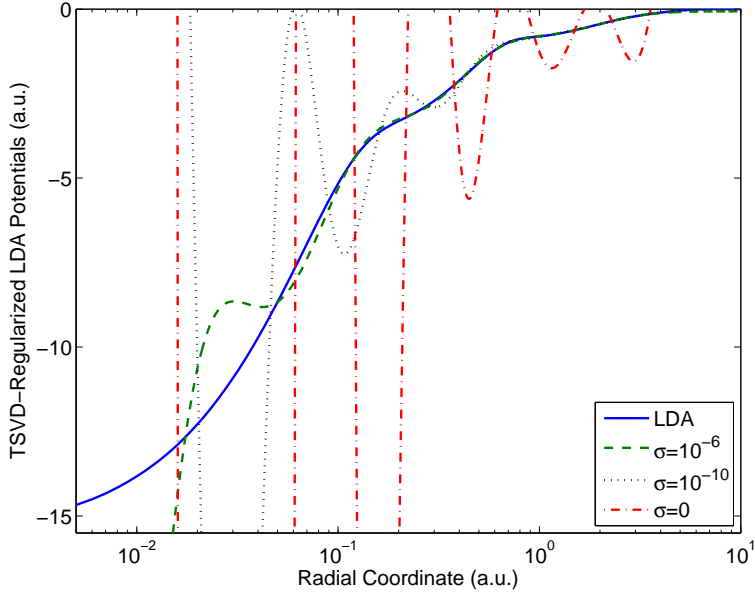}}
\caption{TSVD-regularized LDA potentials of argon for various values
of the SVD cutoff, $\sigma$, using the potential basis Ar8192. The
solid line depicts the exact LDA potential for comparison.}
\end{figure}

The nature of the Hessians spectrum is intimately related to the
behavior of the potentials obtained. Most applications of the direct
OEP procedure have made use of coinciding orbital-quality basis sets
for both orbitals and potential. This, generally, generates physical
potentials and energies. From the Hessian spectrum in Fig. 1 we see
that using the cc-pVDZ orbital and potential basis leads to a full
rank Hessian resulting in the convergence of the OEP to a physical
potential. However, if we use other constructions for the potential
basis such as ET basis sets, we can observe much different spectral
structures for the Hessian. The rank deficiency for the three ET
basis sets is obvious from the significant gap in the spectra of
Fig. 1.  To solve the update equation in an stable manner, both
truncated SVD (TSVD) and Tikhonov regularizations have been proposed
and implemented \cite{WU_YANG_JTCC_03}. The TSVD regularization, to
be used in this discussion, neglects contributions from eigenvectors
corresponding to those singular values below some chosen cutoff.
Fig. 1 shows how the smoothness of the resulting OEP, as measured by
the norm $||\nabla\sum_{t}b_{t}g_{t}(\RR)||$, varies as a function
of the TSVD
cutoff used when solving the Hessian update equation. 
As the cutoff approaches the lower cluster of singular values, the
structure becomes step-like, with jumps in the norm as the cutoff
passes through eigenvalues of the Hessian. 
The potential becomes very nonphysical (Fig. 2), reminiscent of
those displayed in Ref. \cite{staroverov06}. It is clear that this
TSVD regularization, on its own, is not sufficient to ensure a
physically meaningful solution for the OEP in general. It does
however confine irregularities to regions near the nuclei, which
could be sufficient in many applications.




Next, we take the classical approach to ill-posed problems and
further incorporate some desirable measure to regularize the
solution \cite{HANSEN_98}.
It is clear that any nonphysical oscillatory behavior in the
potential will be confined to the basis set expansion
$v_{\bb}=\sum_{t}b_{t}g_{t}$. We thus introduce a
$\lambda$--regularization by
constraining our solutions to yield smooth potentials as measured by
the \emph{smoothing norm}  $||\nabla v_{\bb}||$, thereby restricting
the nonphysical variations in the potential because of unbalanced
 basis set. This norm is certainly not unique, however is
simple to implement and will be seen to produce very satisfying
results. We define a regularized energy functional as
\begin{align}
\Omega_{\lambda}(\bb)=E^{\textrm{YW}}(\bb)+\lambda ||\nabla
v_{\bb}(\RR) ||^{2},
\end{align}
where $||\nabla v_{\bb}(\RR)||^{2} =
2\bb^{\textrm{T}}\textbf{T}\bb$, $\textbf{T}$ is the kinetic energy
integral matrix in the potential basis and $E^{\textrm{YW}}(\bb)$ is
the OEP energy calculated according to the Yang-Wu method. The
energy derivatives with respect to the coefficients are modified
accordingly as
$\nabla_{\bb}\Omega_{\lambda}=\nabla_{\bb}E^{\textrm{YW}}+4\lambda
\textbf{T}\bb$ and
$\nabla^{2}_{\bb}\Omega_{\lambda}=\nabla^{2}_{\bb}E^{\textrm{YW}}+4\lambda
\textbf{T}$.
The significance of the smoothing contribution relative to the
energy minimization is then determined by the magnitude of
$\lambda$. Our method of determining the most appropriate value of
$\lambda$ (that is, the value of $\lambda$  which corresponds to a
physical potential) is the classical L-curve analysis
\cite{HANSEN_98}. This involves constructing the plot of
$\log(||\nabla v_{\bb}||)$ against
$\log(E^{\textrm{YW}}-E^{\textrm{ref}})$ for $\lambda$ over some
appropriate interval. The reference energy, $E^{\textrm{ref}}$, is
taken as $E^{\textrm{HF}}$ for EXX calculations and
$E^{\textrm{LDA}}$ for LDA OEP.

This curve presents the trade-off between the desire to extract a
smooth potential, with that of an unconstrained minimization of the
energy functional which can allow the introduction of nonphysical
potentials. For a \emph{balanced} potential basis set -- defined
here as one that converges to a smooth physical potential --
gradually increasing $\lambda$ from zero has no significant effect
on the norm $||\nabla v_{\bb}||$. In such a situation no
$\lambda$-regularization is needed. On the other hand, a general
 basis can see a significant decrease in the norm as
$\lambda$ increases, until there comes a region where variations in
the the norm subside corresponding to the formation of a stable and
physical potential.

The most appropriate choice of $\lambda$ for any particular physical
application is not always a trivial decision, dependent on the
particular system under consideration and the tolerance that are
acceptable \cite{HANSEN_98}. In the present context, our
calculations have lead to a consistent definition: The $\lambda$
corresponding to the optimal potential, and thus also \emph{the OEP
energy} for the given finite basis sets and chosen smoothing norm,
is that point on the L-curve with minimum slope and, if it is not
unique, that has the minimum value of $||\nabla v_{\bb}||$. Support
for such a choice is presented in \cite{EPAPS}.




Figs. 3 and 5 present such L-curve analysis for our argon example
with LDA and EXX functionals. We use the notation AO/PB$(\lambda)$
to describe calculations performed with AO and PB atomic orbital and
potential basis sets respectively, and with a regularization
parameter of $\lambda$. For consistency we chose to perform the
calculations without any TSVD regularization. The curve for
cc-pVDZ/cc-pVDZ $(\lambda)$ shows that no regularization needs to be
applied to solve for an appropriate OEP as explained previously,
however the cost of using such a basis is an  energy higher than
that can be obtained from more flexible ET basis sets. The most
appropriate LDA OEP potential for this basis set (cc-pVDZ/cc-pVDZ
$(0)$) as seen in Fig. 4 underestimates the LDA
potential near the nucleus.


\begin{figure}[t]
\centering{\includegraphics[height=8cm]{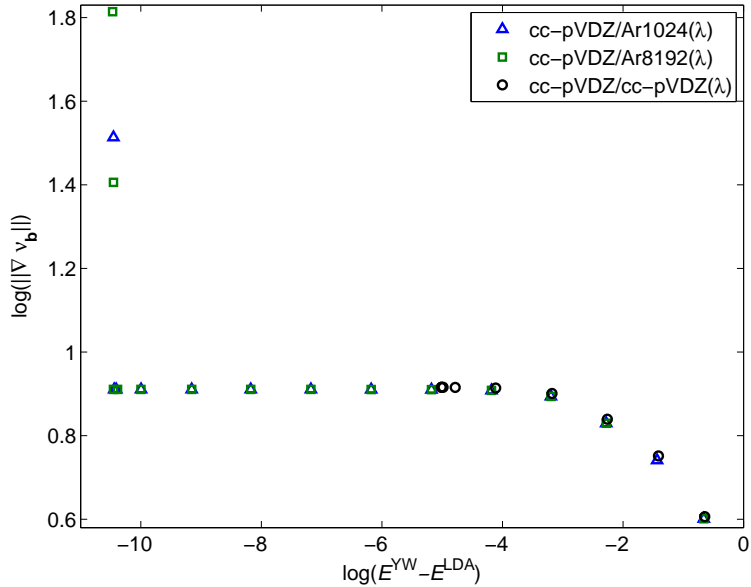}}
\caption{Argon (cc-pVDZ) LDA OEP L-curve for various potential basis
sets.}
\end{figure}

\begin{figure}[t]
\centering{\includegraphics[height=8cm]{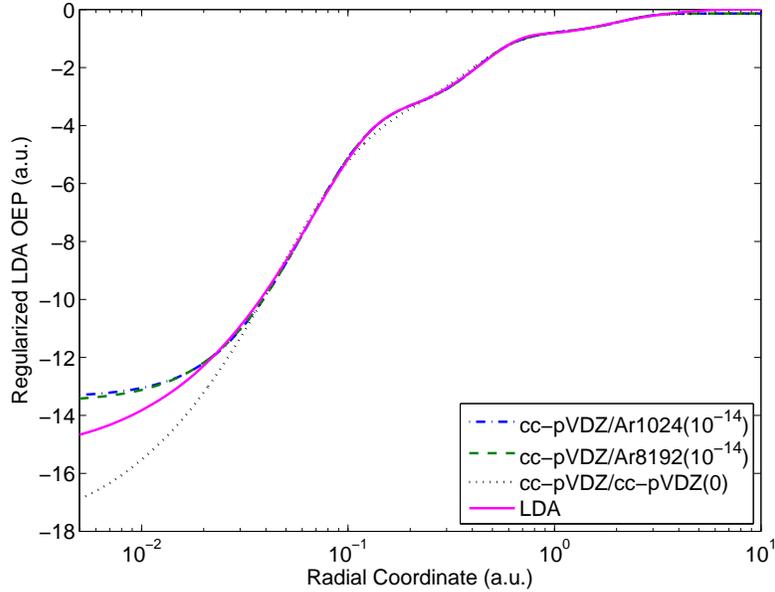}}
\caption{Optimal LDA OEP potentials for argon (cc-pVDZ) obtained
from the L-curve analysis for each basis set.}
\end{figure}

All three of the ET basis sets produce poor potentials with
oscillations extending far from the nuclei when no regularization is
used. The LDA L-curves for cc-pVDZ/Ar8192$(\lambda)$ and
cc-pVDZ/Ar1024$(\lambda)$ are of an ideal form -- as $\lambda$ tends
towards zero smooth potentials are generated with numerical energy
agreement with the LDA value, at which point the curve rises
vertically corresponding to the introduction of nonphysical
oscillations in the potential. In this situation the choice of a
meaningful potential is clear, the corner of the L-curve is well
defined and $\lambda^*=10^{-14}$, the optimal choice of $\lambda$,
is that point infinitesimally close to the corner (with minimum
slope). The corresponding potential (Fig. 4) agrees with the LDA
potential in all but a small region around the nuclei.




The EXX results for the same basis sets in Figs. 5 and 6 show that
the critical value of $\lambda$ occurs at $\sim\Delta E=10^{-4}$
a.u. Significantly, this example illustrates that  nonphysical
potentials  can be obtained when the ill-posedness is not addressed,
and that the physical OEP potentials can be obtained with our
method. As another example, the EXX OEP L-curve analysis for
$\textrm{N}_{2}$ is shown in Fig. 7. We see that all basis sets
regularize remarkably consistently, giving rise to a consistent OEP
energy that will be explored in some detail within a future paper.
Further detailed examples for atoms and molecules are provided
online via EPAPS \cite{EPAPS}.




\begin{figure}[t]
\centering{\includegraphics[height=8cm]{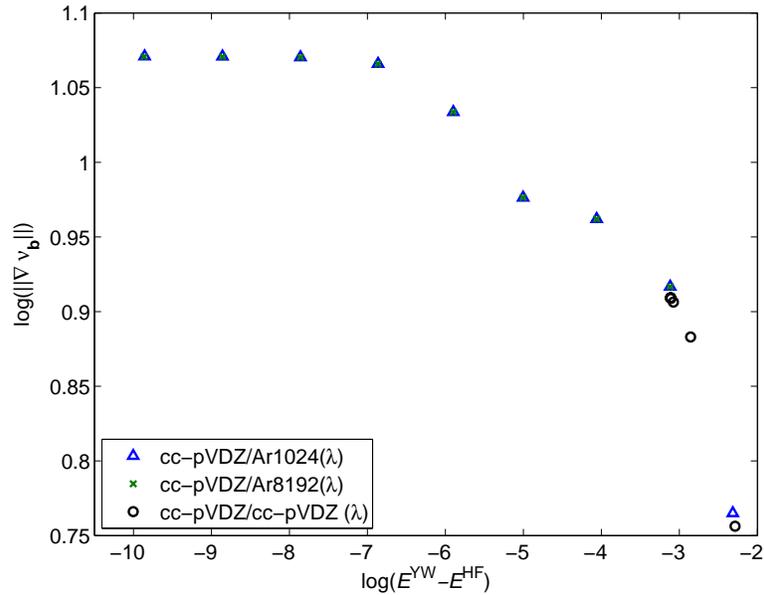}}
\caption{Argon (cc-pVDZ) EXX L-curve for various potential basis
sets.}
\end{figure}

\begin{figure}[t]
\centering{\includegraphics[height=8cm]{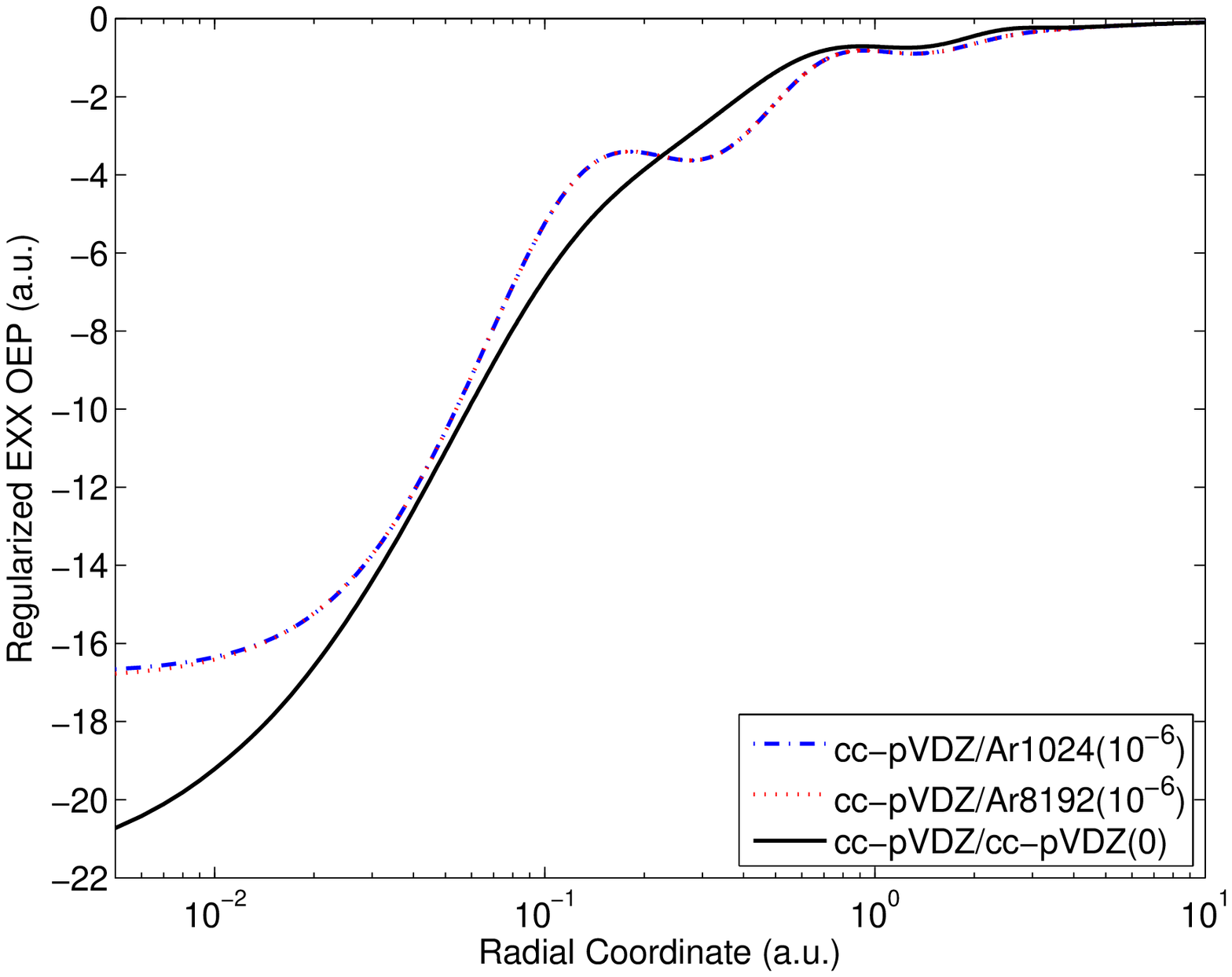}}
\caption{Optimal EXX OEP potentials for argon (cc-pVDZ) obtained
from the L-curve analysis for each basis set.}
\end{figure}

\begin{figure}[t]
\centering{\includegraphics[height=8cm]{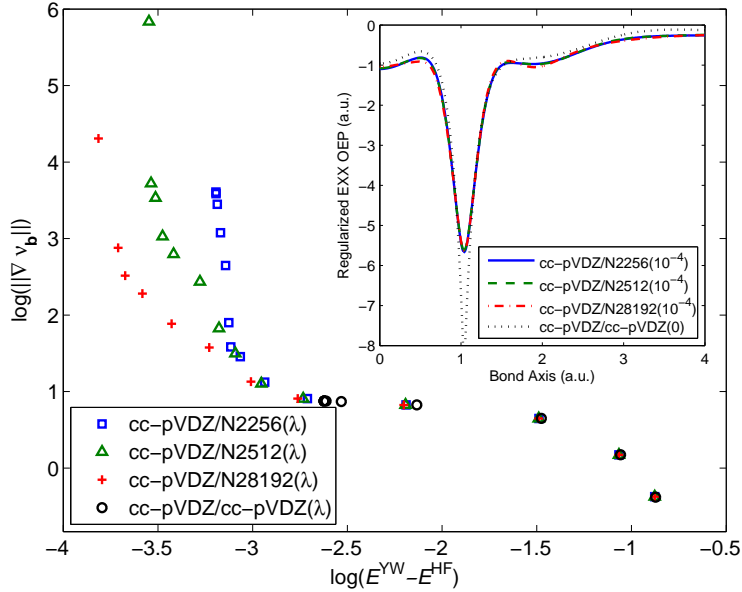}} \caption{EXX
L-curve of $\textrm{N}_{2}$ (cc-pVDZ) with four potential basis
sets. The basis sets are defined in \cite{EPAPS}. Insert: optimal
potentials as determined by the L-curve.}
\end{figure}

In conclusion, we have constructed a regularization procedure by
introducing a measure of the smoothness of the OEP into the energy
functional. This, together with an L-curve analysis, allows us to
determine the physically meaningful OEP potential and energy from
calculations performed with any finite basis sets. We have shown
that one quantitative measure of such balance is provided by the
approximate Hessian, however this does not provide an a priori
judgement needed for routine application of finite basis OEP
calculations. For this, the construction of balanced potential basis
sets are required.
Support from the National Science Foundation is gratefully
acknowledged.


\newpage

\begin{center}\Huge{\textbf{Supplementary Material}}\end{center}

\section{Our choice of the optimal regularization parameter $\lambda^{*}$}

Firstly, we wish to consider our choice of $\lambda^{*}$: The value
of the regularization parameter corresponding to the optimal choice
of potential. In many statistical applications, it is recommended to
choose that corner of the L-curve with maximum curvature, for this
point is argued to represent the optimal tradeoff between stability
of solutions, and absolute minimization of the quantity of interest
\cite{HANSEN_98}. This however is dependent on the physical system
under consideration and the tolerances acceptable to the user. We
find the maximum curvature point on the L-curve not to be optimal
for our concern -- irregularities have been introduced and the
regularized potential has began to diverge away from the physical
potential. The over-smoothing at the corner of the L curve  has also
been observed in some applications \cite{HANSEN_98}.

\begin{figure}[h]
\centering{\includegraphics[height=8cm]{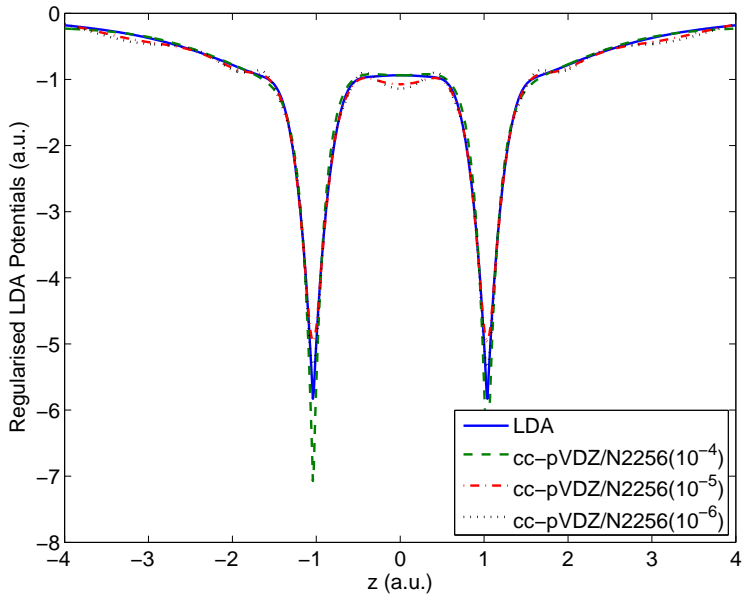}}
\caption{LDA OEP potentials for $\textrm{N}_{2}$ (cc-pVDZ) in the
vicinity of the L-curve corner for the potential basis
N2256.}\label{esp:pot256_lda_n2}
\end{figure}

The following example shows, for an LDA calculation of nitrogen
(potential basis set is introduced in section III), the changes in
the potential as we traverse the corner of the L-curve in Fig.
(\ref{esp:pot256_lda_n2}). $\lambda=10^{-4}$ best corresponds to
that point on the L-curve with minimum slope, (or minimum absolute
gradient), while the maximum curvature corresponds to
$\lambda=10^{-6}$. The irregularities appearing with the different
$\lambda$ are most noticeable at the origin and for large $z$.

In the following we will go though several more examples showing the
versatility of our procedure.

\section{Water}

We shall consider the orbital basis of cc-pVDZ, and the three
potential basis sets cc-pVDZ, H2O01, and H2O02. H2O01 and H2O02 are
even tempered (ET) basis sets constructed as follows: H2O01 -- 9s9p
(O) and 9s (H) uncontracted cartesian Gaussians with exponents
$2^{n}, -3<n<5$. H2O02 -- 18s15p (O) and 18s (H) uncontracted
cartesian Gaussians with exponents $2^{n}, -4<n<13$ (s) and $2^{n},
-4<n<13$ (p). Our construction of the ET basis sets are such to
encourage poor behavior in the resulting, unregularized, potentials
so to show the versatility of our procedure.

We use the geometry 0.917{\AA} for the O-H bond length and 104.4
degrees for the HOH angle, and plot potentials along the HOH
bisector.

\subsection{LDA}

\begin{figure}[htb]
\centering{\includegraphics[height=8cm]{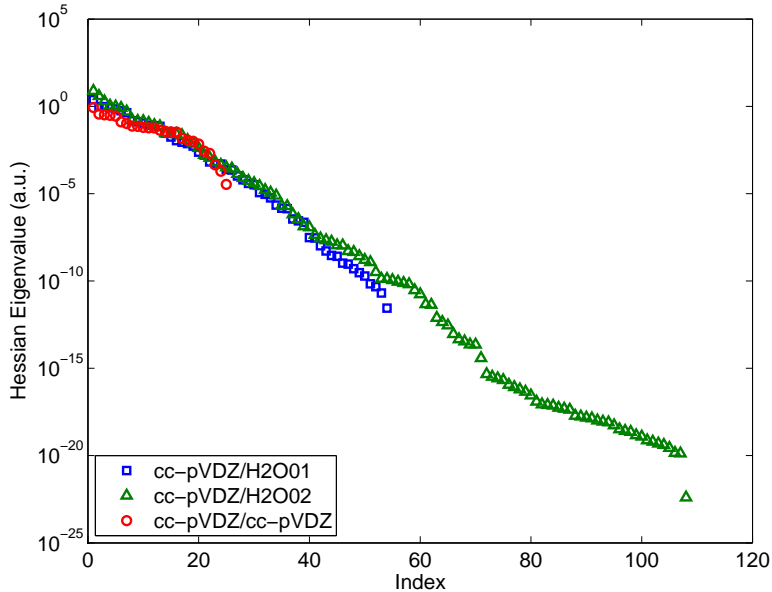}}
\caption{Spectrum of the approximate Hessian for each basis set.}
\label{esp:hessian_lda_water}
\end{figure}

For water, the Hessian spectra are shown in Fig.
(\ref{esp:hessian_lda_water}). As is typical for molecules, it
becomes much harder to interpret the Hessian spectra in terms of a
definable cluster of singular values that we would consider to be
associated with null-space eigenvectors. The problem becomes very
much ill-conditioned for the basis H2O02: the eigenvectors decay
gradually to (numerically) zero, as opposed to the situation
presented in the paper for Ar where the problem is very much rank
deficient.

\begin{figure}[htb]
\centering{\includegraphics[height=8cm]{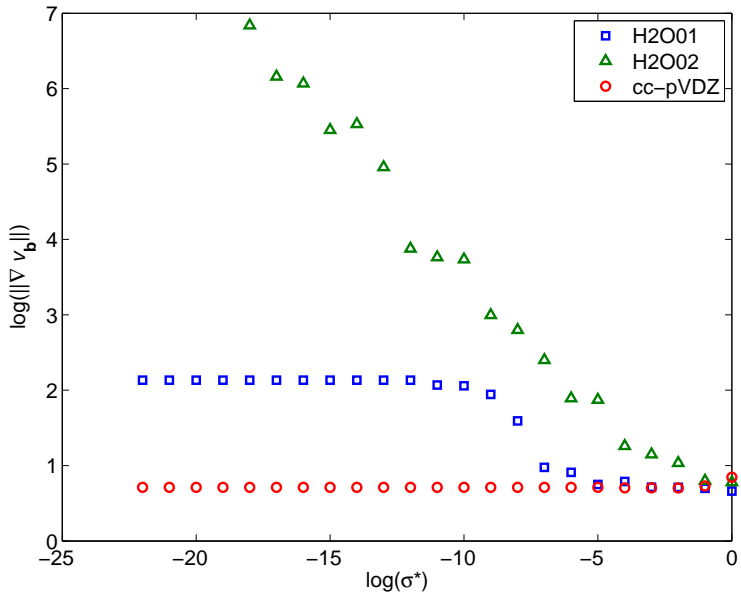}}
\caption{TSVD regularization} \label{esp:svdlshape_lda_water}
\end{figure}

\begin{figure}[htb]
\centering{\includegraphics[height=8cm]{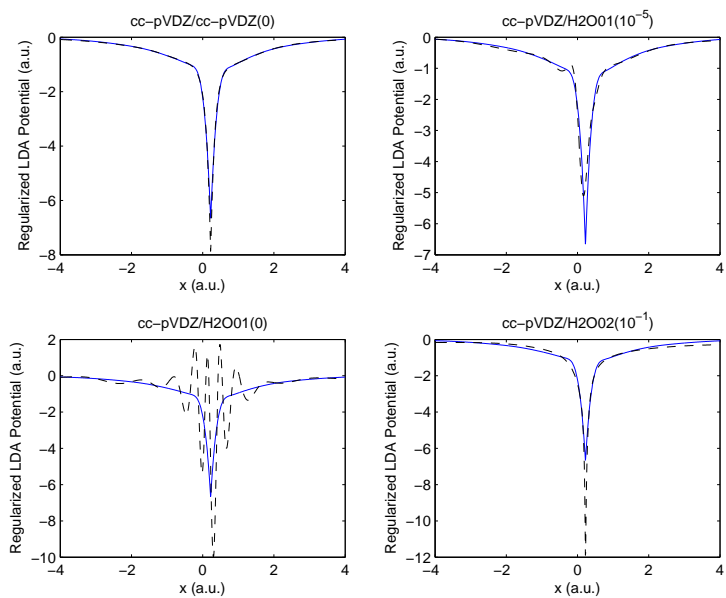}}
\caption{TSVD-regularized LDA potentials for water (cc-pVDZ) for
various values of the cutoff. The solid line depicts the exact LDA
potential.} \label{esp:potentials_lda_water}
\end{figure}

The variation of $||\nabla v_{\textbf{b}}||$ as a function off the
TSVD cutoff, $\sigma^{*}$, is presented in Fig.
(\ref{esp:svdlshape_lda_water}), in an attempt for extract a
definable range of eigenvalues for which we can consider the Hessian
to be stable if used in solving the update equation
$\textbf{H}(\textbf{b}_{n})\textbf{p}=-\nabla E(\textbf{b}_{n})$.
The step structure as was seen in the main text is still present,
however for the H2O02 basis it becomes very hard to argue that any
particular value of the TSVD cutoff would be any more favorable for
a TSVD-only regularized solution for the OEP. The other two basis
sets provide very clear interpretations in terms of the
eigenspectrum of Fig. (\ref{esp:hessian_lda_water}). H2O01 has a
smallest eigenvalue of only $3.5\times 10^{-12}$, so that any choice
for a TSVD cutoff smaller than this will yield identical results.
The resulting potential is, however, very poor with nonphysical
oscillations about the oxygen atom as seen in Fig.
(\ref{esp:potentials_lda_water}). Choosing a value for the cutoff
around $10^{-5}$ does lead to much better potentials, but still
there are some irregularities around the nucleus. The situation for
cc-pVDZ/cc-pVDZ is typical, the Hessian eigenvalues decay to only
$\sim 3\times 10^{-5}$ a.u.

\begin{figure}[htb]
\centering{\includegraphics[height=8cm]{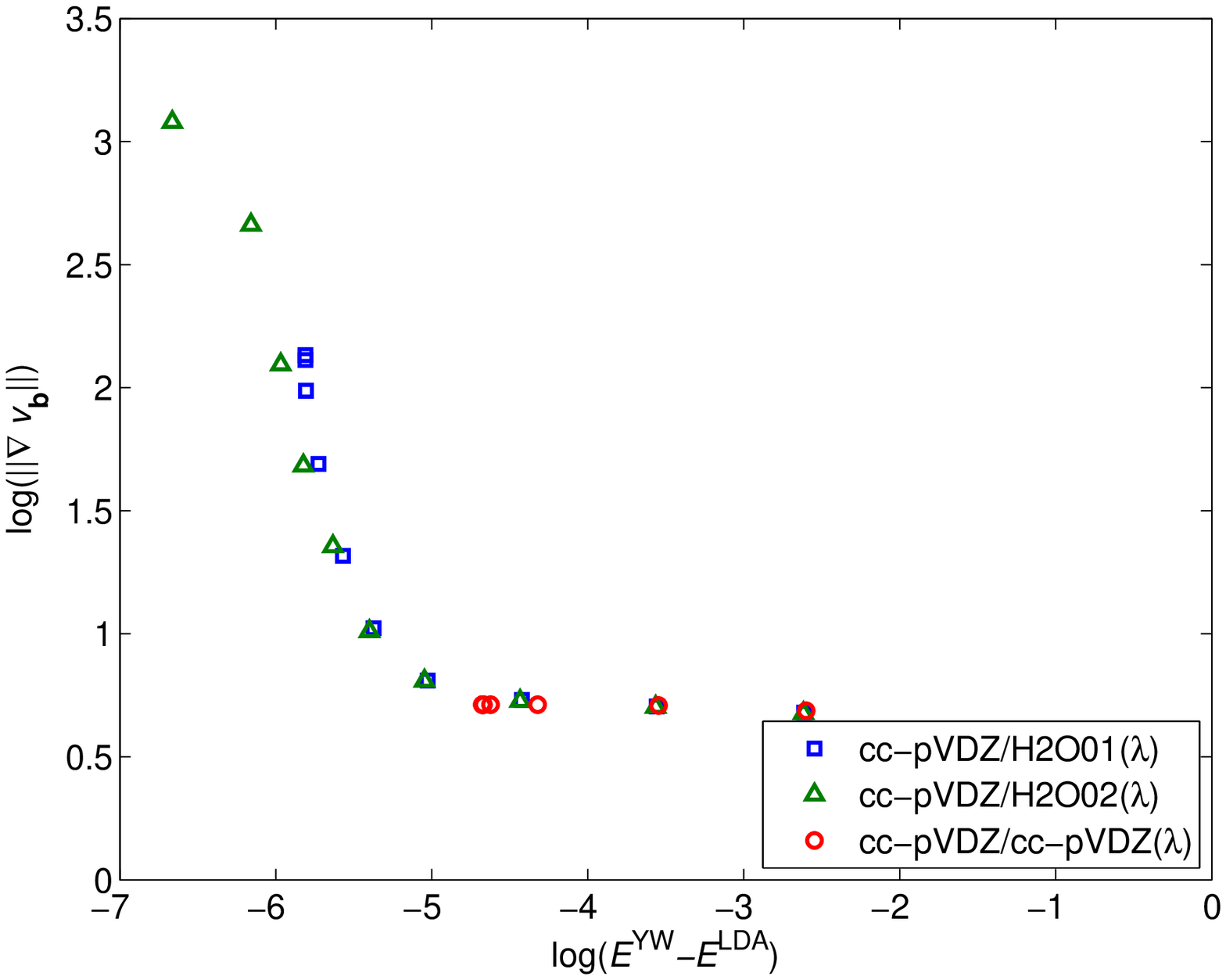}}
\caption{LDA L-curves for water in the various potential basis
sets.}\label{esp:lshape_lda_water}
\end{figure}

\begin{figure}[htb]
\centering{\includegraphics[height=8cm]{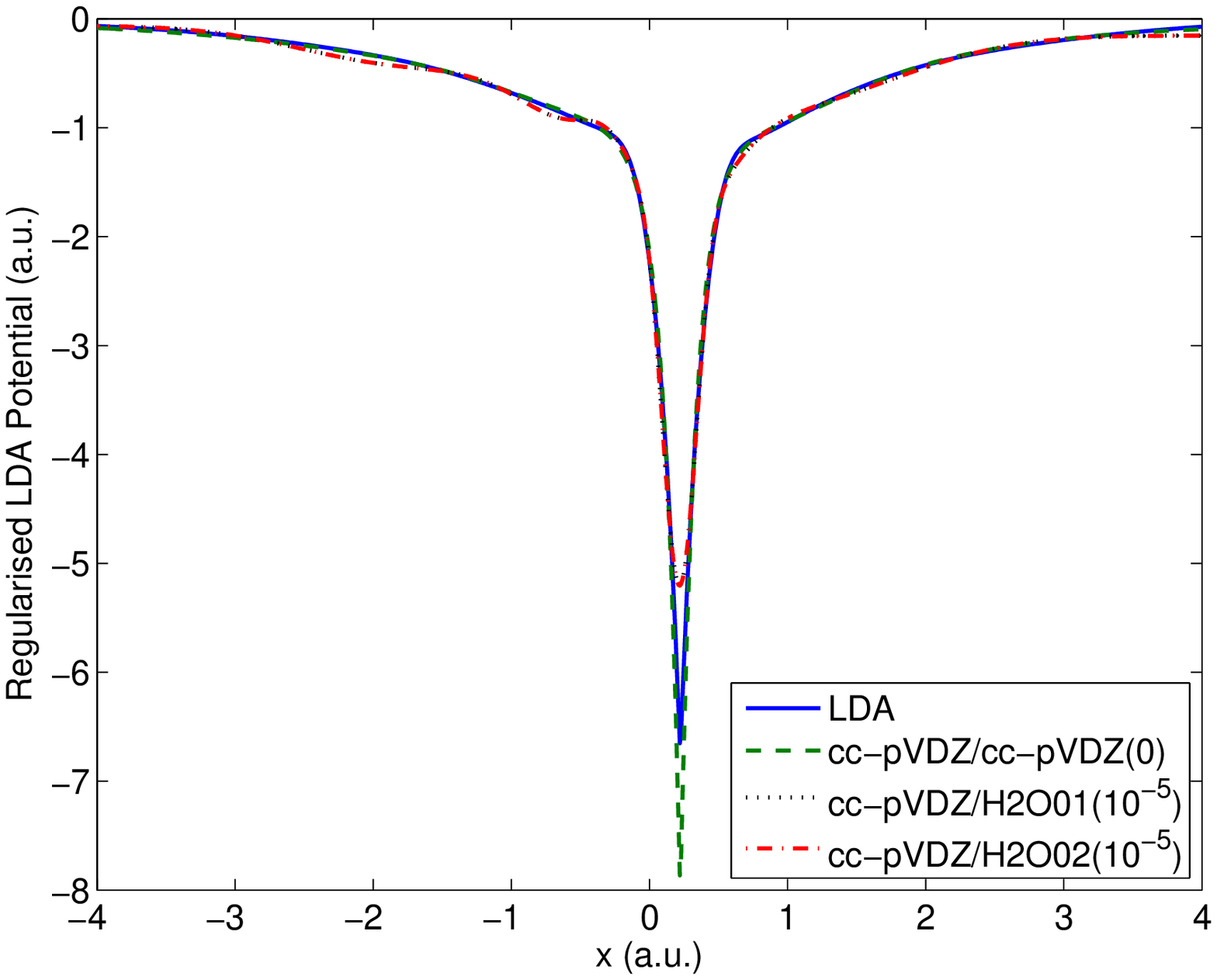}}
\caption{Optimal LDA OEP potentials for water (cc-pVDZ) obtained
from the L-curve analysis for each basis
set.}\label{esp:pot_lda_water}
\end{figure}

It is clearly necessary to apply the $\lambda$-regularization for
both ET basis sets in order to produce appropriate potentials. We
chose not to apply any TSVD when solving the update equation and
obtain the L-curves seen in Fig. (\ref{esp:lshape_lda_water}), in
turn giving the potentials, corresponding to the minimum slope
point, in Fig. (\ref{esp:pot_lda_water})

\subsection{EXX}

The Hessian structure is very similar to that of the LDA case. This
is a very general observation for all the systems we have examined,
so the discussion for the LDA case carries over to EXX. As such we
will specifically only discuss the L-curves and the potentials we
can extract in the EXX case although a complete set of figures are
presented.

\begin{figure}[htb]
\centering{\includegraphics[height=8cm]{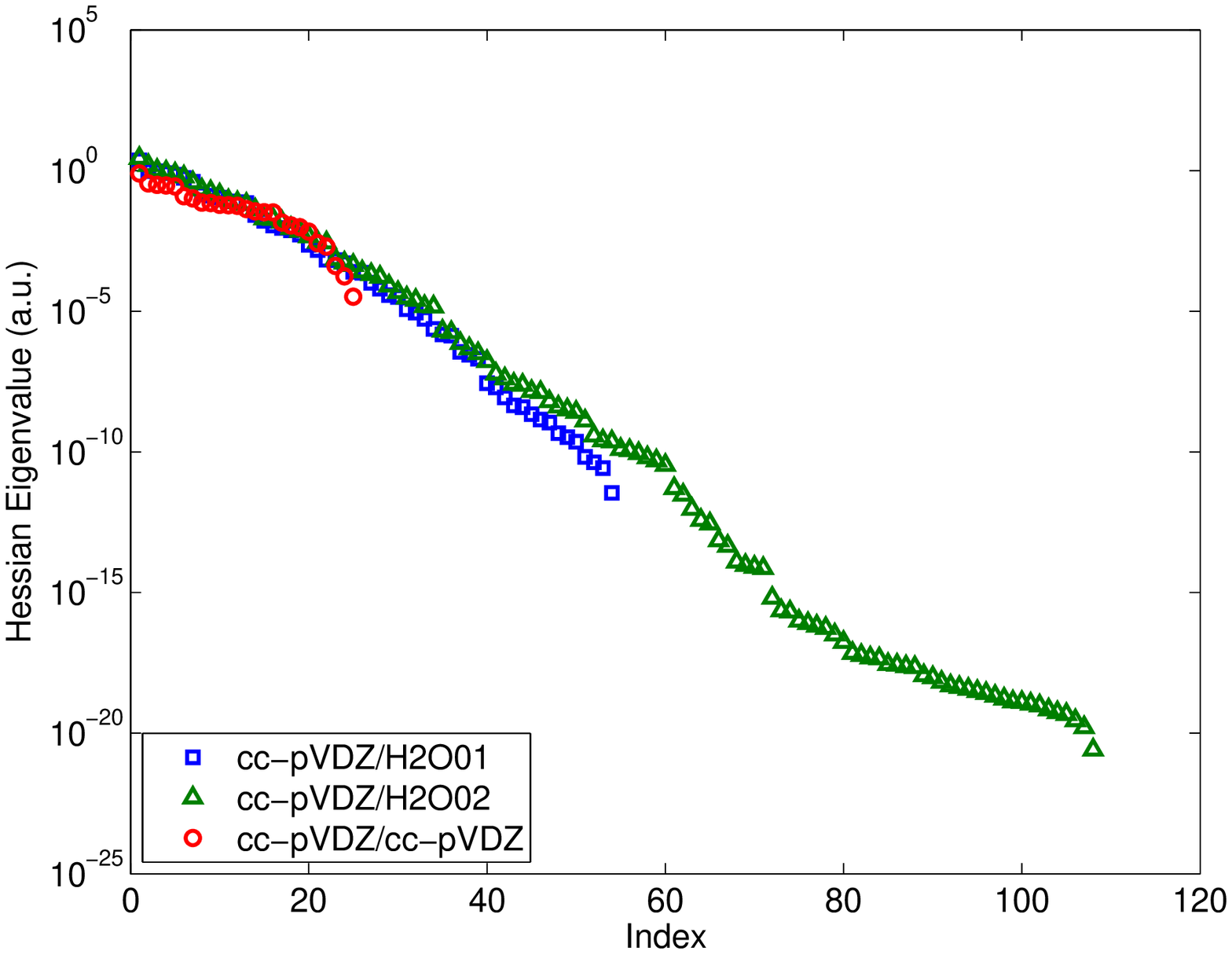}}
\caption{Spectrum of the approximate Hessian for each basis set.}
\label{esp:hessian_exx_water}
\end{figure}
\begin{figure}[htb]
\centering{\includegraphics[height=8cm]{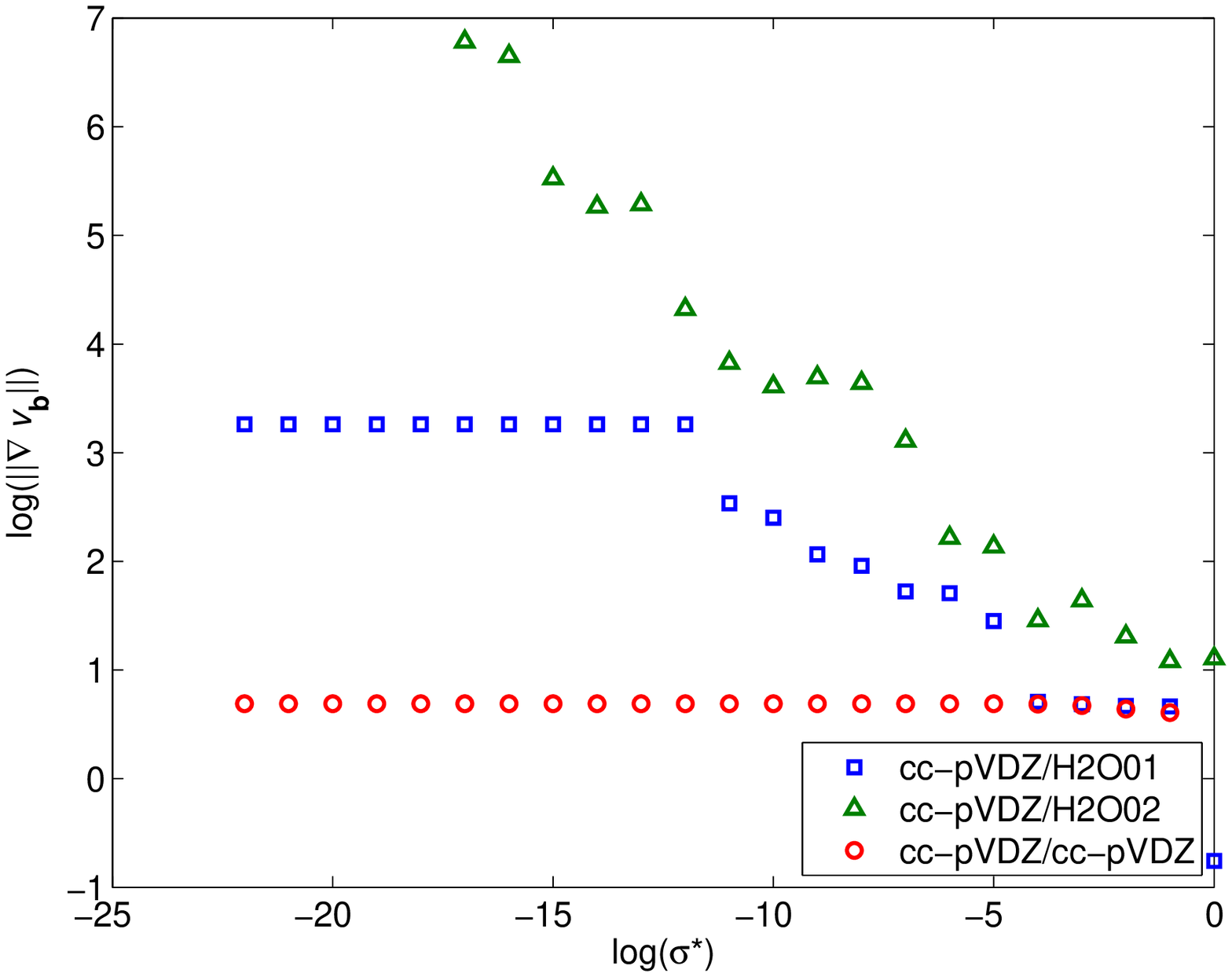}}
\caption{TSVD regularization} \label{esp:svdlshape_exx_water}
\end{figure}
\begin{figure}[htb]
\centering{\includegraphics[height=8cm]{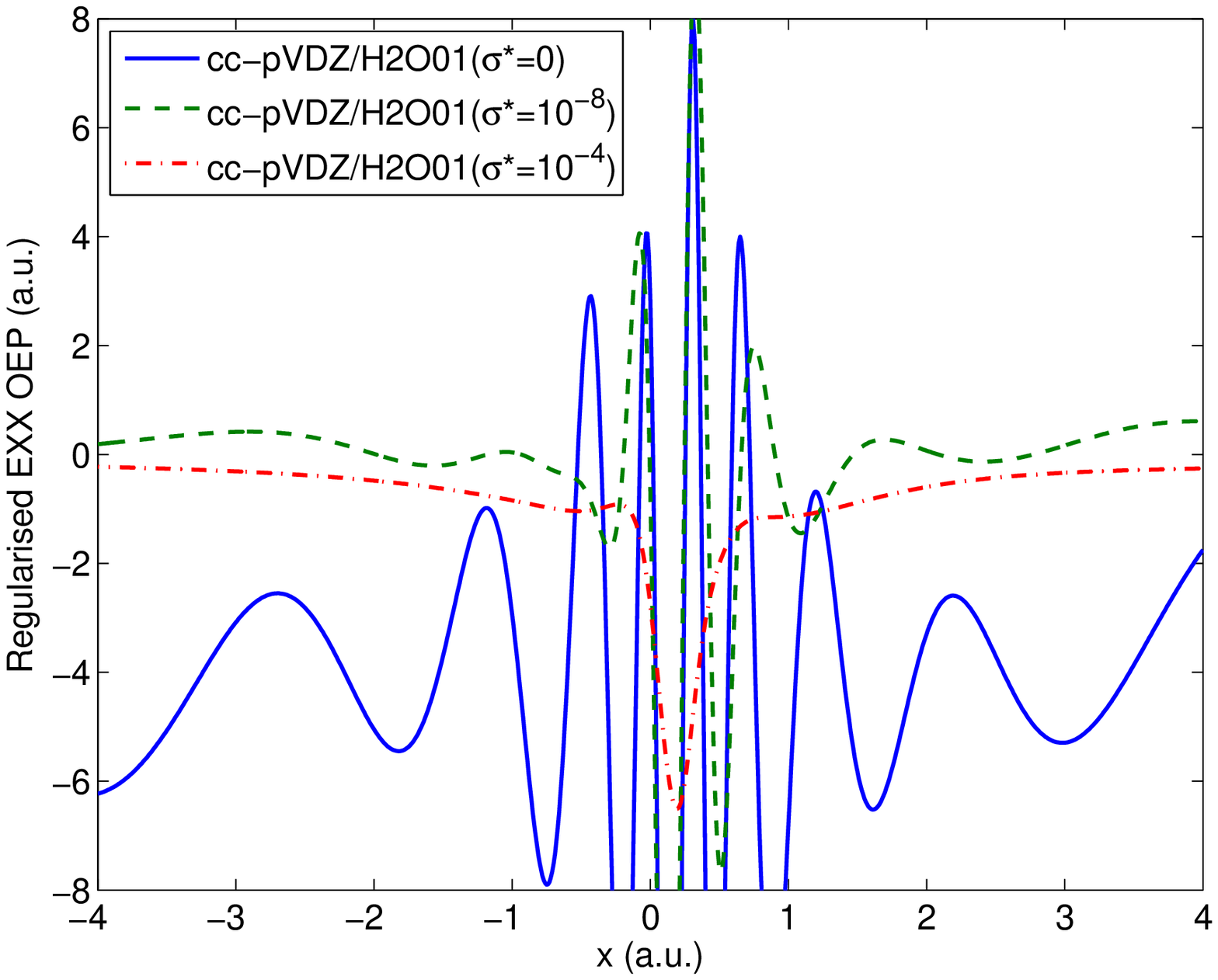}}
\caption{TSVD-regularized EXX potentials for water (cc-pVDZ/H2O01)
for various values of the TSVD cutoff}
\end{figure}

\begin{figure}[htb]
\centering{\includegraphics[height=8cm]{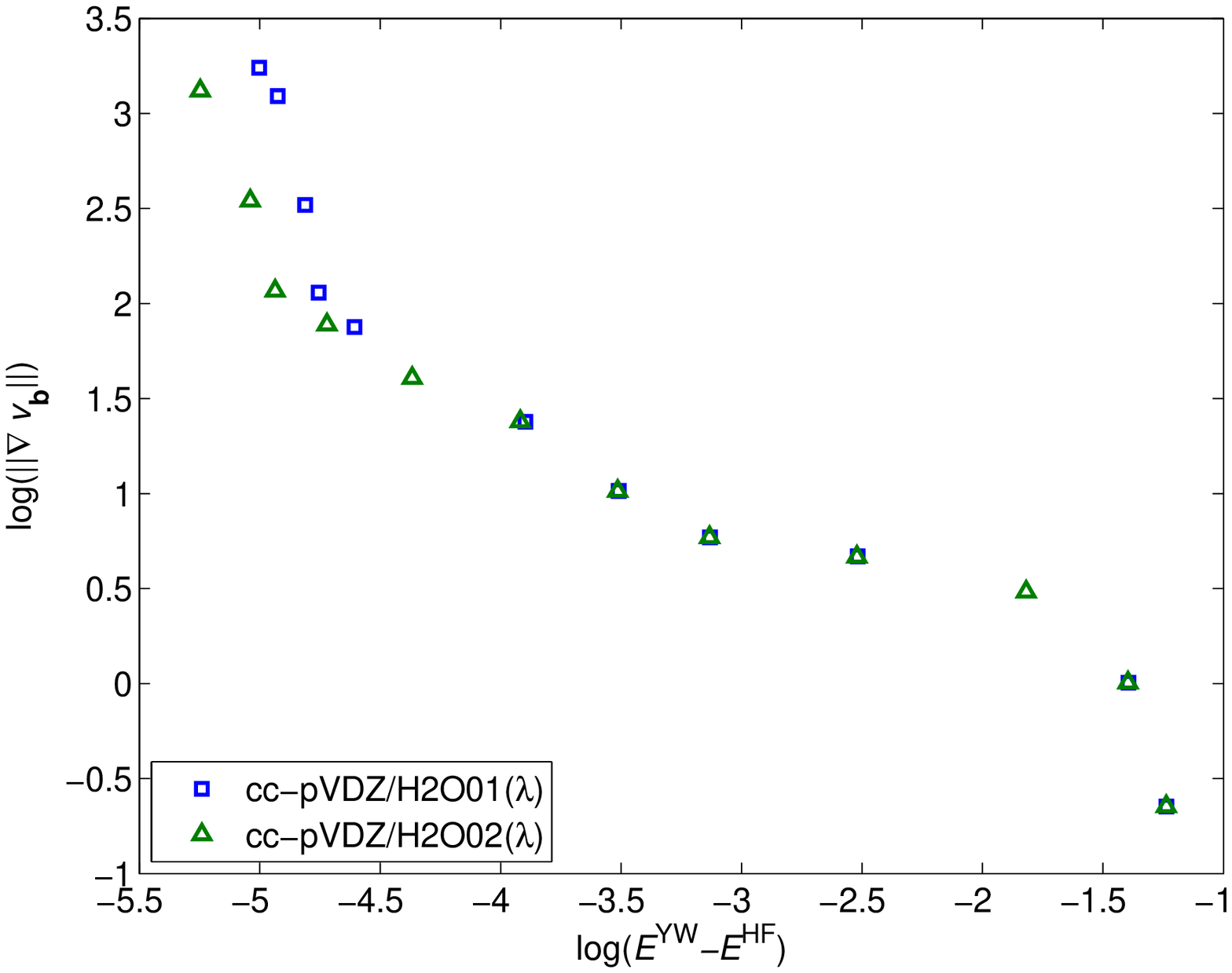}}
\caption{LDA L-curves for water in the various potential basis
sets.}\label{esp:lshape_exx_water}
\end{figure}

\begin{figure}[htb]
\centering{\includegraphics[height=8cm]{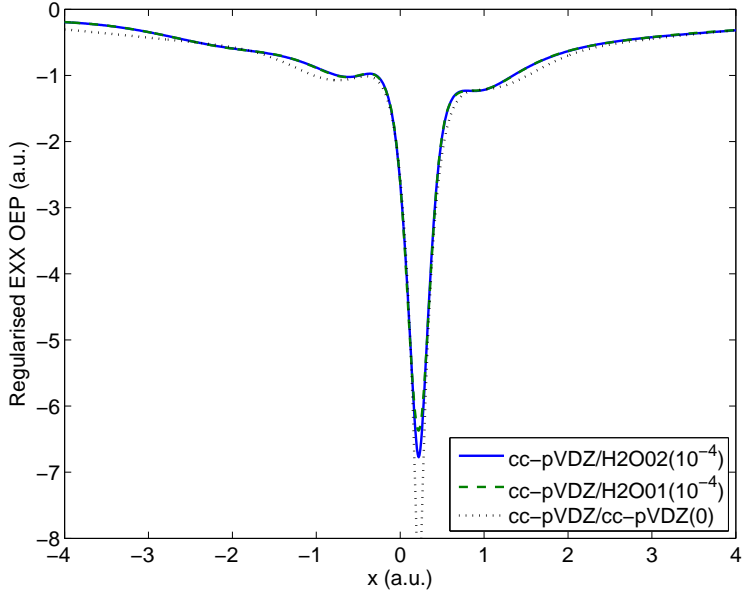}}
\caption{Optimal EXX OEP potentials for water (cc-pVDZ) obtained
from the L-curve analysis for each basis
set.}\label{esp:pot_exx_water}
\end{figure}

The L-curves of Fig. (\ref{esp:lshape_exx_water}) agree
exceptionally well for all but the smallest amounts of
regularization. This is very encouraging given the differences in
the potential basis sets. An inflexion point is well defined and
gives an approximation to the OEP energy of
$E^{\textrm{OEP}}-E^{\textrm{HF}} \sim 1 \times 10^{-2.5}$ a.u.. If
we want to get more accurate energies a finer $\lambda$-sampling
grid would be necessary. The given basis sets can reproduce the HF
energy to an agreement of $\sim 10^{-5}$ a.u. with nonphysical
potentials. The potentials we extract from the L-curve are shown in
Fig. (\ref{esp:pot_exx_water}). The two ET basis sets produce
essentially identical potentials, differing only at the oxygen cusp.
The potential cc-pVDZ/cc-pVDZ(0) is also of an acceptable quality,
but as expected with the use of basis sets designed for orbitals, it
does struggle to reproduce the finer details that a flexible ET
basis can.

\section{Nitrogen}

The potential basis sets we use for $\textrm{N}_{2}$ are as denoted
as N2256, N2512, and N28192,  consisting of 13s3p, 14s5p and 18s8p
uncontracted cartesian Gaussians with exponents given in Table
\ref{exponents_n2}. An orbital basis if cc-pVDZ is used. We use the
a bond length of 1.098{\AA}, and plot potentials along the bond
axis.

\begin{table}[htbp]
\begin{tabular}{c|c||c}
    &   s-exponents   &   p-exponents  \\
\hline
N2256 (13s3p) & $2^{n}, -4\leq n\leq 8$ &   $2^{n}, -2\leq n\leq 0$\\
N2512 (14s5p) & $2^{n}, -4\leq n\leq 9$ &   $2^{n}, -3\leq n\leq 1$\\
N28192 (18s8p) & $2^{n}, -4\leq n\leq 13$ &   $2^{n}, -2\leq n\leq 5$\\
\end{tabular}
\label{exponents_n2} \caption{Exponents used in constructing the
potential basis sets used for our $\textrm{N}_{2}$ calculations.}
\end{table}

\subsection{LDA}

The results for nitrogen are very much similar to those for water
previously discussed. All three ET basis sets are capable of
producing very oscillatory potentials. The L-curve analysis in Fig.
(\ref{esp:lshape_lda_n2}) provides appropriate potentials (Fig.
(\ref{esp:pot_lda_n2})).

\begin{figure}[htb]
\centering{\includegraphics[height=8cm]{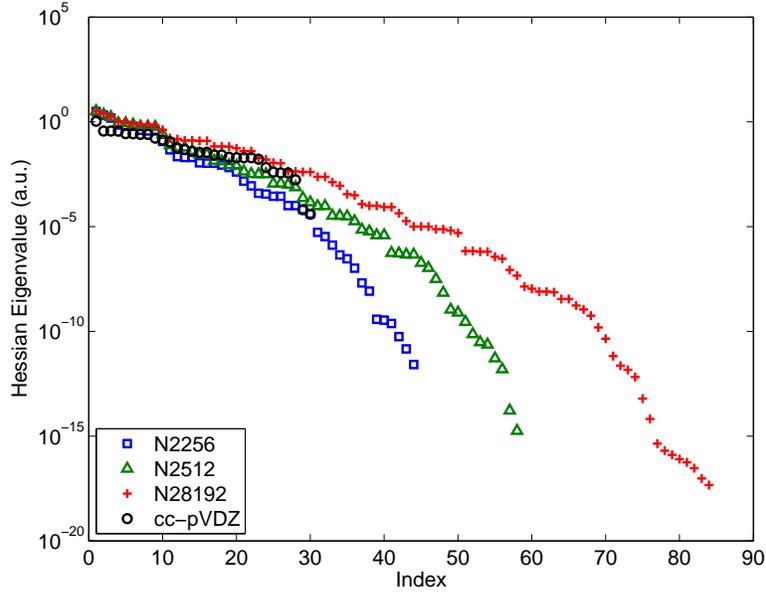}}
\caption{Spectrum of the approximate Hessian for each basis set.}
\label{esp:hessian_lda_n2}
\end{figure}

\begin{figure}[h]
\centering{\includegraphics[height=8cm]{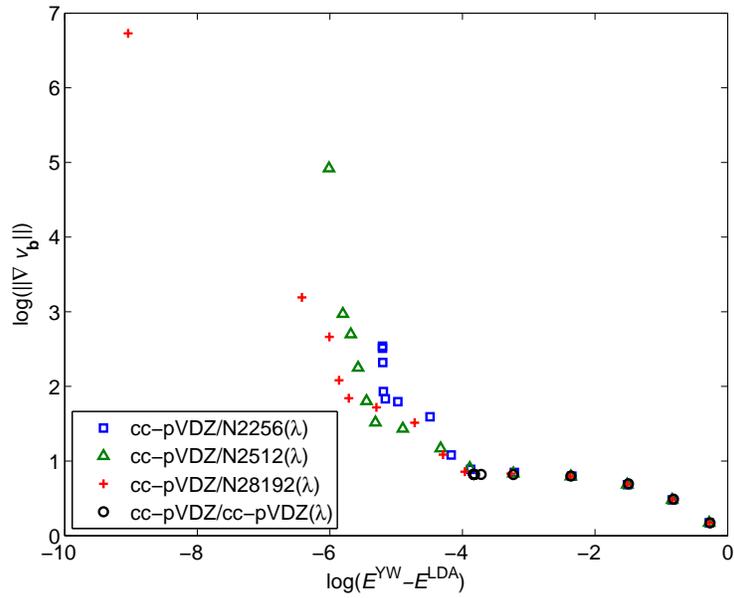}}
\caption{$\lambda$ regularization}\label{esp:lshape_lda_n2}
\end{figure}

\begin{figure}[h]
\centering{\includegraphics[height=8cm]{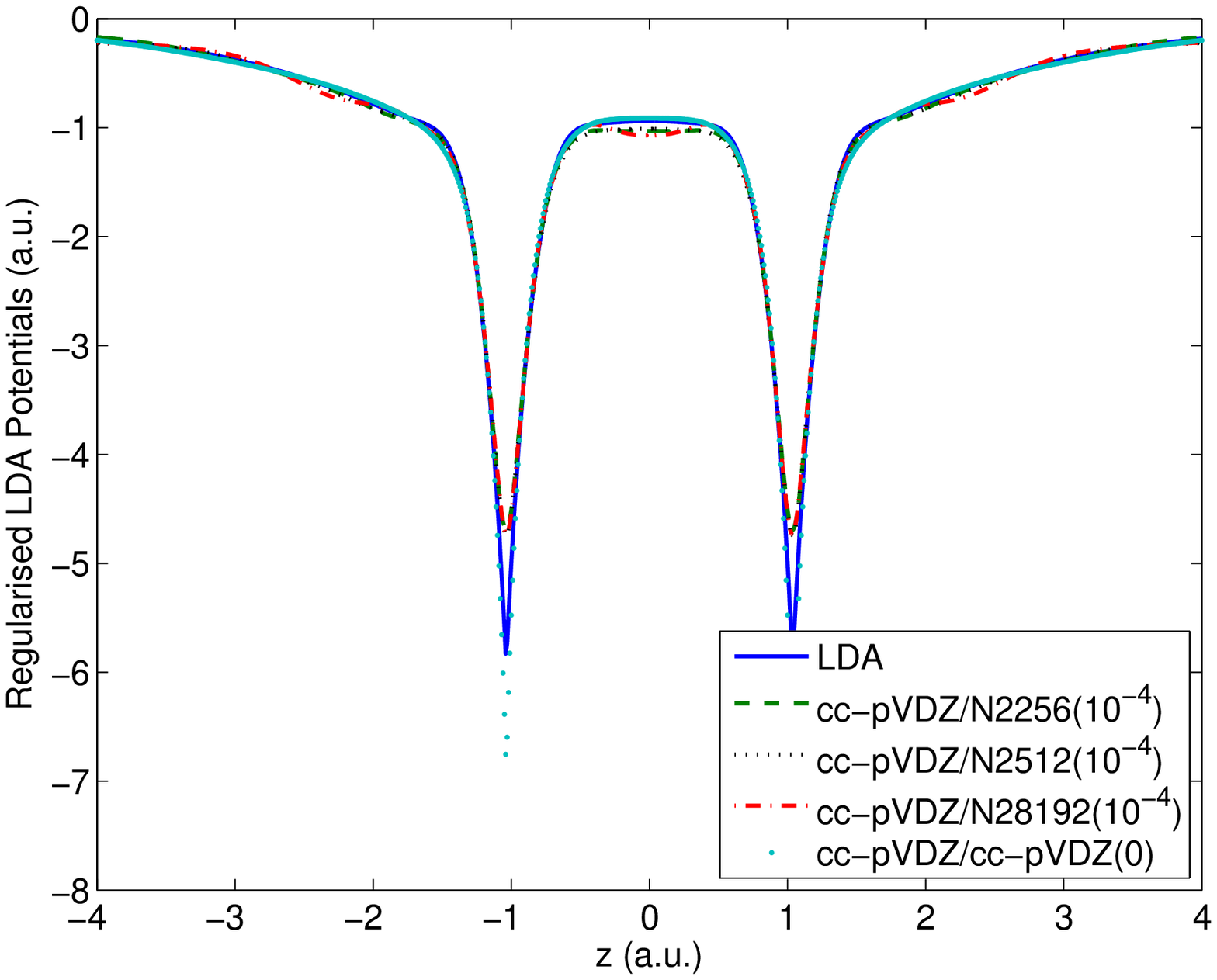}}
\caption{Optimal LDA OEP potentials for $\textrm{N}_{2}$ (cc-pVDZ)
obtained from the L-curve analysis for each basis
set.}\label{esp:pot_lda_n2}
\end{figure}


\subsection{EXX}

The L-curve in Fig. (\ref{esp:lshape_exx_n2}) and potentials have
been presented in the text for this case. Included here is the
Hessian spectrum, and examples of $\lambda$-regularized potentials
for non-optimal $\lambda$.

\begin{figure}[h]
\centering{\includegraphics[height=8cm]{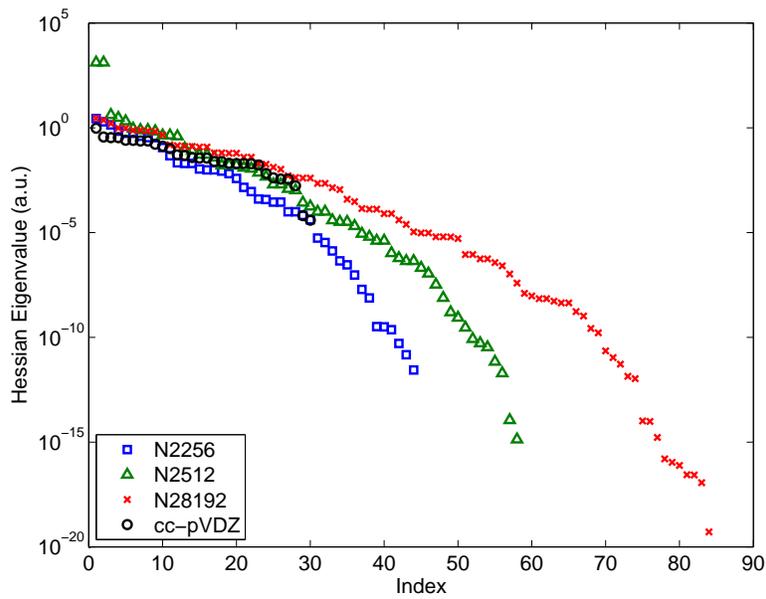}}
\caption{Spectrum of the Approximate Hessian for each basis set.}
\label{esp:hessian_exx_n2}
\end{figure}

\begin{figure}[h]
\centering{\includegraphics[height=8cm]{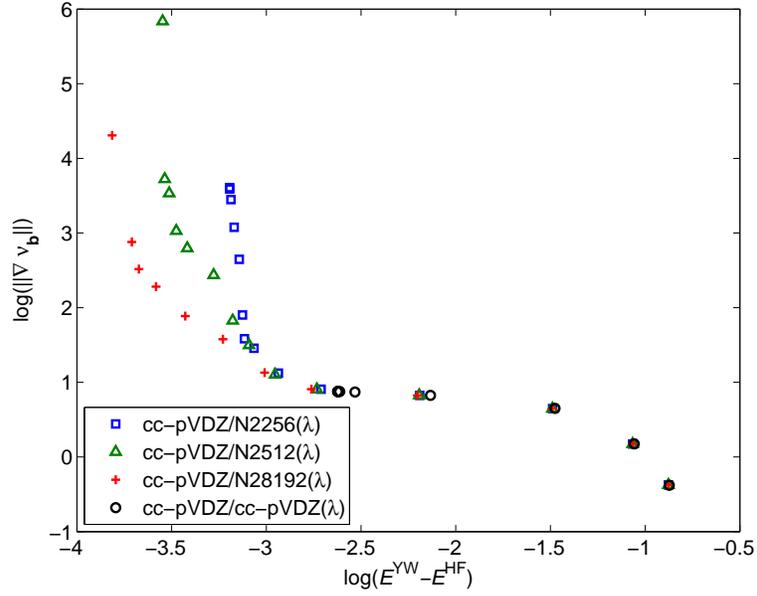}}
\caption{EXX L-curves for $\textrm{N}_{2}$ in the various potential
basis sets.}\label{esp:lshape_exx_n2}
\end{figure}

\begin{figure}[h]
\centering{\includegraphics[height=8cm]{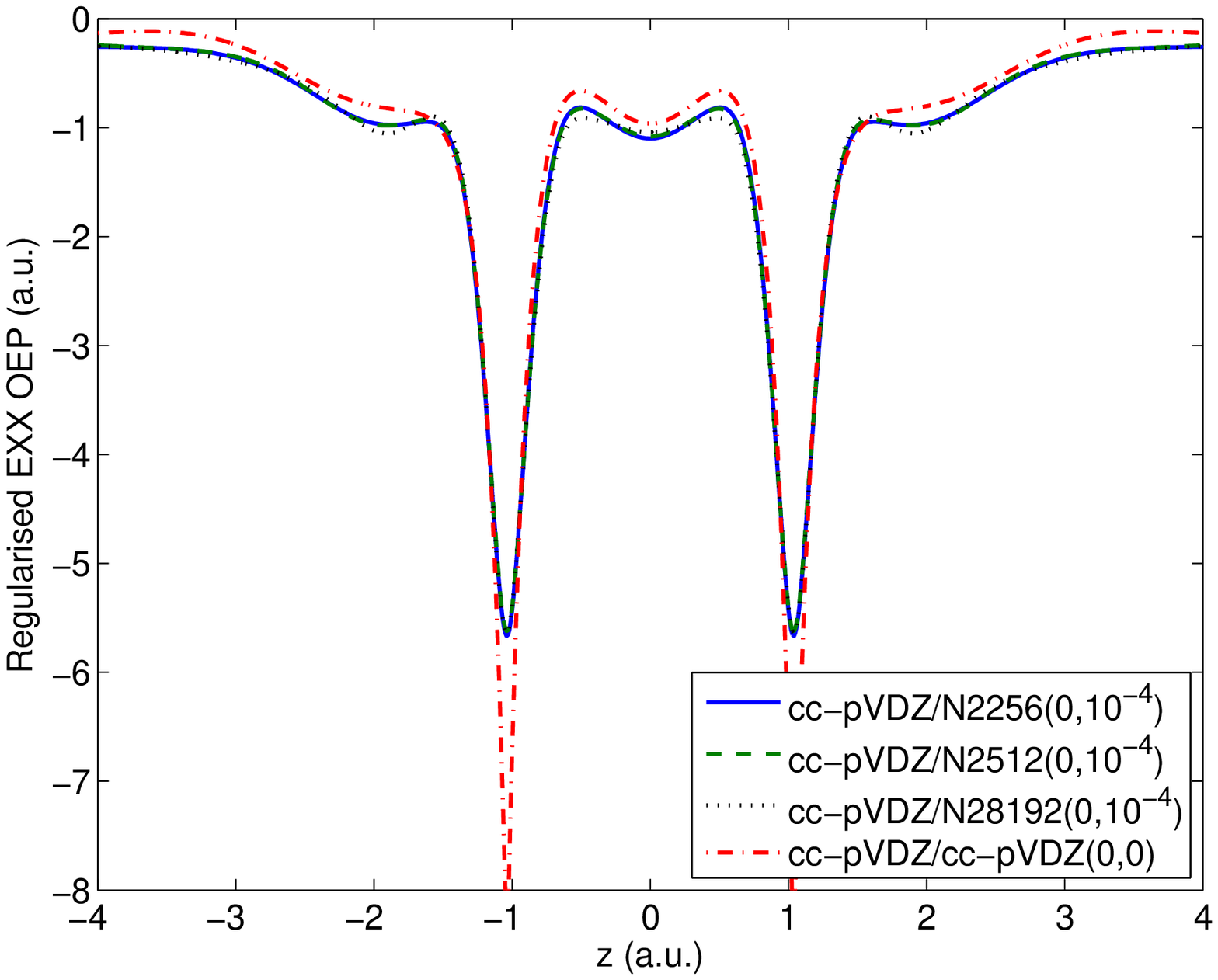}}
\caption{Optimal EXX OEP potentials for $\textrm{N}_{2}$ (cc-pVDZ)
obtained from the L-curve analysis for each basis set.}
\end{figure}

\begin{figure}[h]
\centering{\includegraphics[height=8cm]{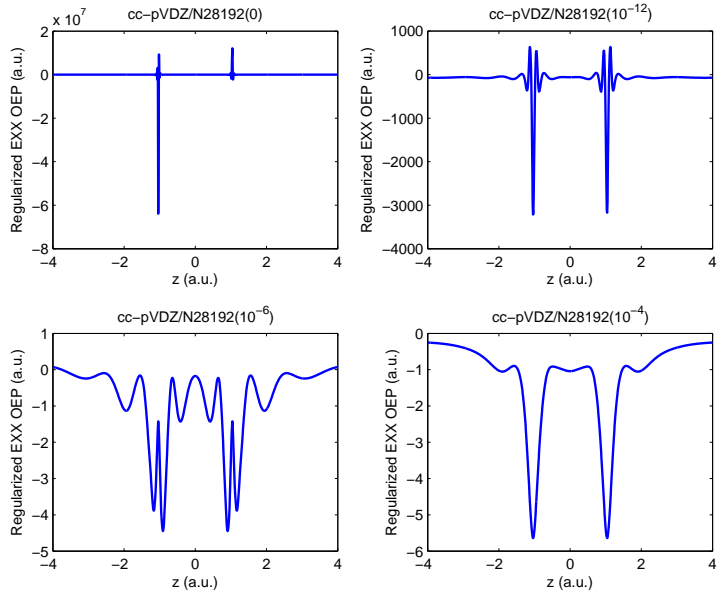}}
\caption{$\lambda$-regularized potentials from various points on the
L-curve for the calculation cc-pVDZ/N2256}
\end{figure}

\section{Argon}

\subsection{EXX}

Here we look at the effects of using a very high quality basis set,
the Partridge Uncontracted 3 basis. For the potential basis set we
use the Partridge Uncontracted 3 set itself, as well as a 20s ET
basis Ar32768 with exponents $2^{n},-4\leq n \leq 15$.

\begin{figure}[htb]
\centering{\includegraphics[height=8cm]{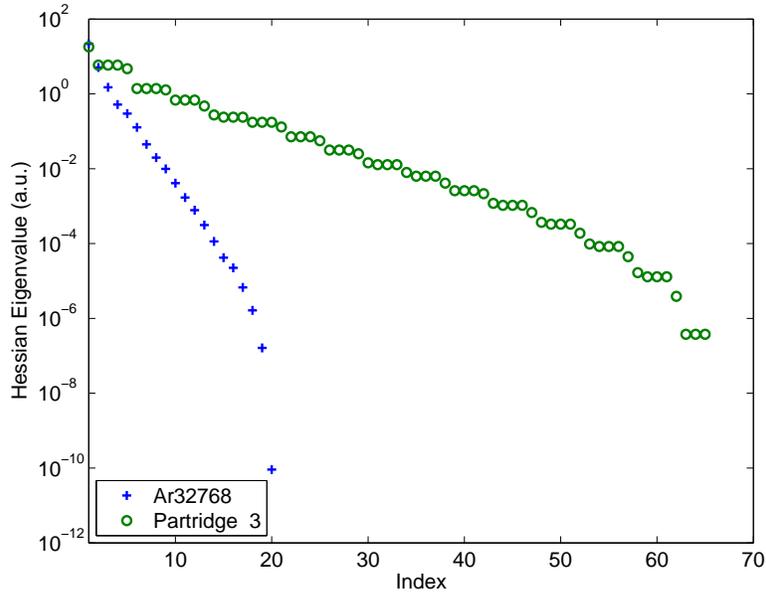}}
\caption{Spectrum of the Approximate Hessian for each basis set.}
\label{esp:hessian_pu3 ar}
\end{figure}

\begin{figure}[htb]
\centering{\includegraphics[height=8cm]{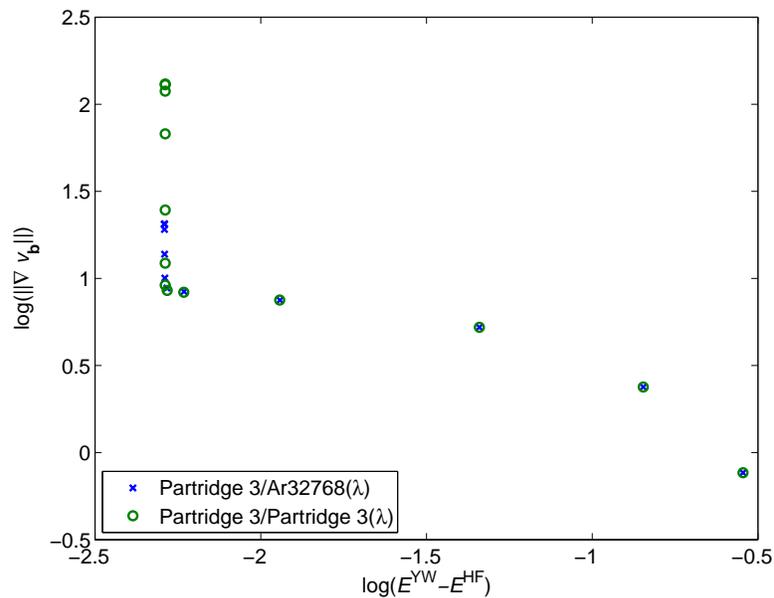}}
\caption{EXX L-curves for Ar (Partridge Uncontracted 3)
}\label{esp:lshape_pu3_ar}
\end{figure}

The L-curve of Fig. (\ref{esp:lshape_pu3_ar}) is interesting as it
shows an example where the use of coinciding basis sets for orbitals
and potential does give rise to nonphysical potentials. This would
be difficult to predict on the basis of the Hessian spectrum of Fig.
\ref{esp:hessian_pu3 ar} alone.

Also notable from the L-curve is the very well defined EXX OEP
energy $\Delta E = 5.2\times 10^{-3}$ a.u.

\begin{figure}[htb]
\centering{\includegraphics[height=8cm]{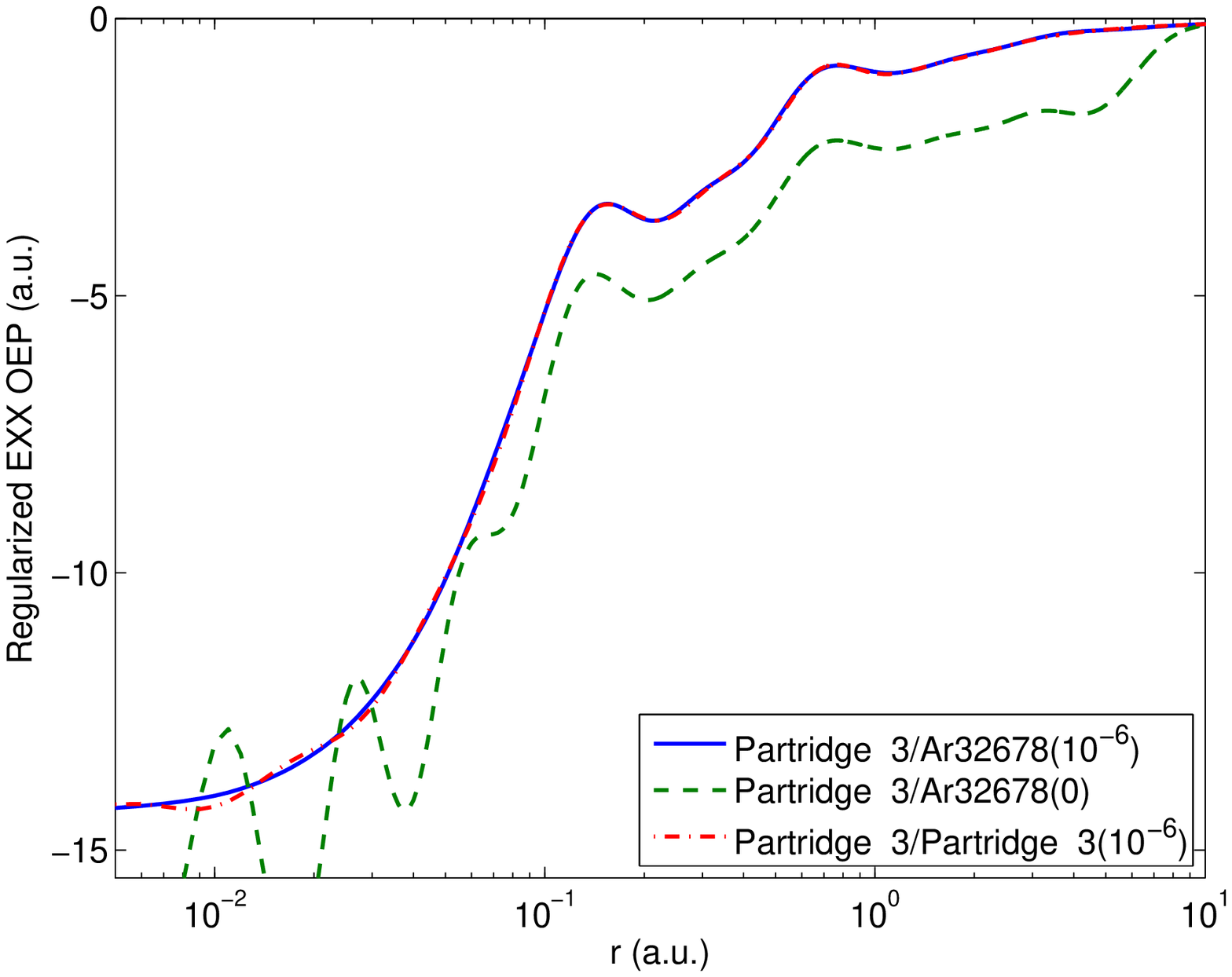}}
\caption{Optimal regularized potentials for each basis set, and an
unregularised potential}
\end{figure}

\subsection{Small Potential Basis Sets}

\begin{figure}[htb]
\centering{\includegraphics[height=8cm]{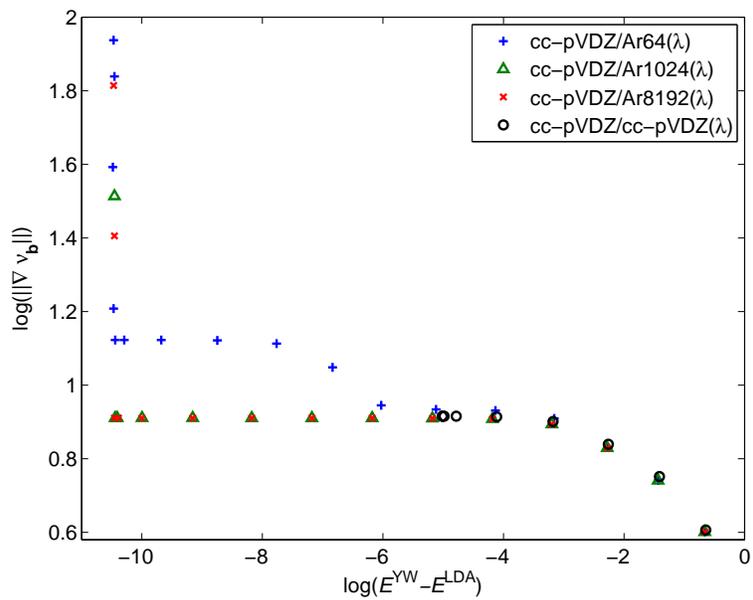}}
\caption{LDA L-curves for Ar (cc-pVDZ)}
\label{esp:arall_lda_lshape_1}
\end{figure}

\begin{figure}[htb]
\centering{\includegraphics[height=8cm]{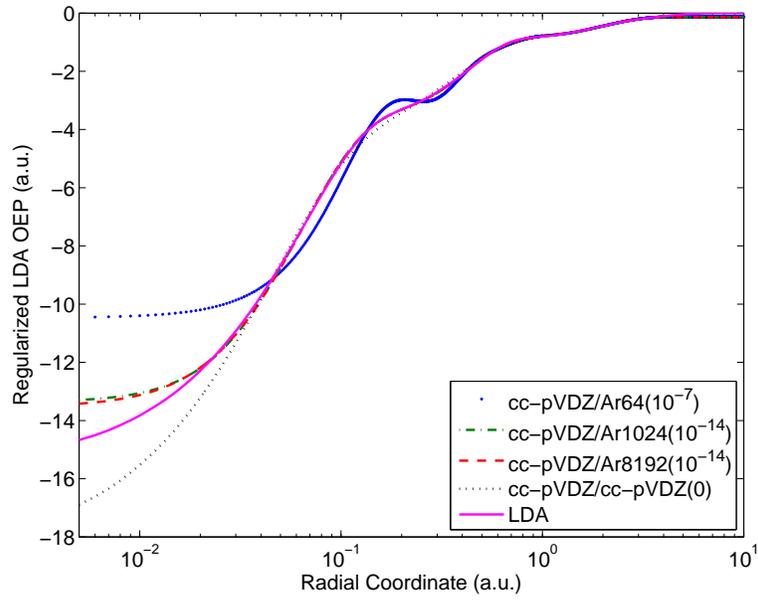}}
\caption{Optimal regularized LDA potentials for each basis set}
\label{esp:lda_optpot_1}
\end{figure}

\begin{figure}[htb]
\centering{\includegraphics[height=8cm]{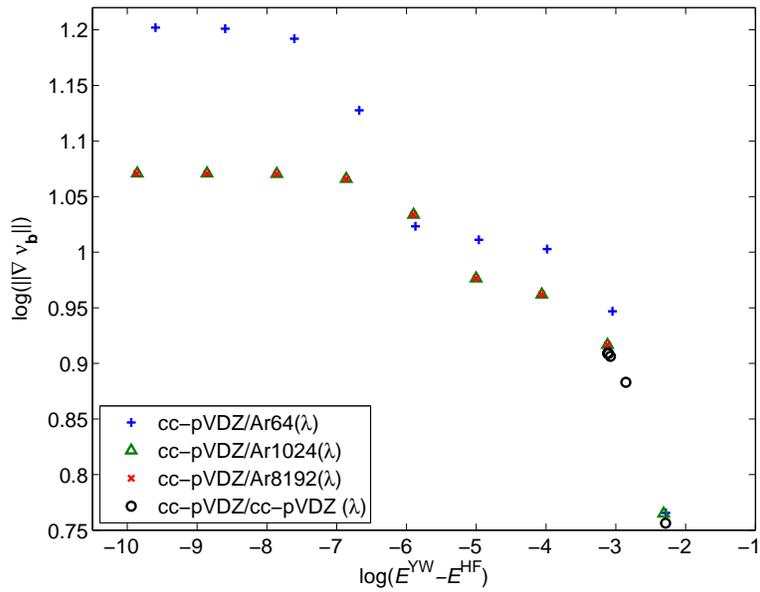}}
\caption{EXX L-curves for Ar (cc-pVDZ)}
\label{esp:arall_exx_lshape_1}
\end{figure}

\begin{figure}[htb]
\centering{\includegraphics[height=8cm]{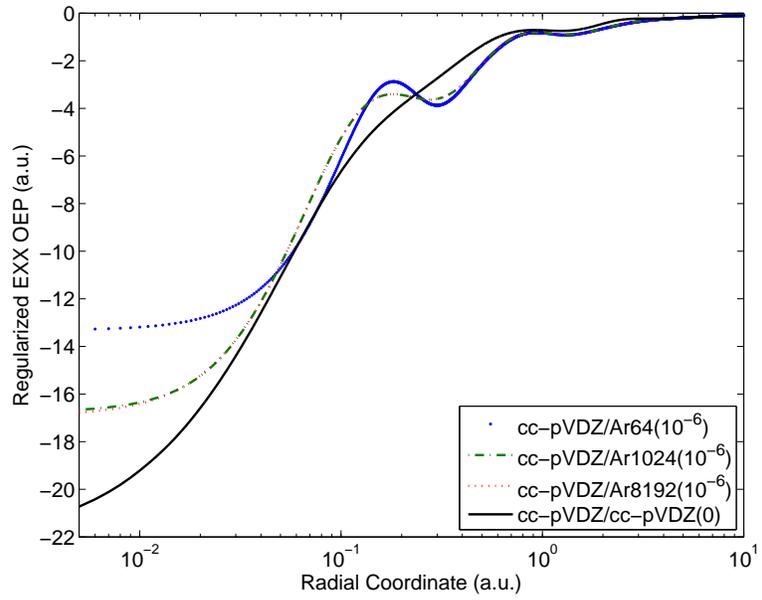}}
\caption{Optimal regularized EXX potentials for each basis set}
\label{esp:ar_exx_svdregpot_1}
\end{figure}

Finally, we consider the situation where we use a basis set that is
so small and inflexible that it is not able to provide an adequate
description of the OEP. The final three figures repeat those shown
in the paper for argon, however now with the inclusion of the small
potential basis set Ar64. While this basis set gives significantly
different L-curves (Figs. (\ref{esp:arall_lda_lshape_1}) and
(\ref{esp:arall_exx_lshape_1}) for LDA and EXX respectively) and
potentials (Figs. (\ref{esp:lda_optpot_1}) and
(\ref{esp:ar_exx_svdregpot_1})) from those larger basis sets, we see
that the L-curve analysis still is able to generate a sensible
potential differing only near the nuclei.

\end{document}